\documentclass[prb,aps,preprint]{revtex4}
\usepackage{graphicx,amsmath}

\begin{document}

\title{Electron transport through dipyrimidinyl-diphenyl diblock molecular wire: protonation effect}

\author{Zhenyu Li\footnote{E-Mail: zhenyuli@umd.edu.}}

\address{Department of Chemistry and Biochemistry,
 University of Maryland, College Park, 20742}


\begin{abstract}
Recently, rectifying direction inversion has been observed in
dipyrimidinyl-diphenyl (PMPH) diblock molecular wire [J. Am. Chem.
Soc. (2005) 127, 10456], and a protonation mechanism was suggested
to explain this interesting phenomena. In this paper, we study the
protonation effect on transport properties of PMPH molecule by first
principles calculations. No significant rectification is found for
the pristine diblock molecular wire. Protonation leads to
conductance enhancement and rectification. However, for all
considered junctions with rectifying effect, the preferential
current directions are samely from dipyrimidinyl side to diphenyl
side. Effect of molecule-electrode anchoring geometry is studied,
and it is not responsible for the discrepancy between experiment and
theory.
\end{abstract}

\maketitle

\section{Introduction}

Molecular rectifier is the first example proposed for electronic
device at molecular scale, \cite{aviram7477} and it is an
important topic in molecular electronics. \cite{nitzan0384,
joachim0501} Considerable experimental and computational efforts
have been devoted in recent years to molecular rectification.
\cite{Metzger0303} In a recent experiment on rectification of
dipyrimidinyl-diphenyl (PMPH) diblock molecular wire,
\cite{Morales0556} the preferential current direction was found to
be able to change possibly by protonation of the dipyrimidinyl
moiety. In this experiment, PMPH molecule was first co-assembled
on the Au(111) substrate together with dodecane through thiol
group. Then the top end of PMPH molecule was connected to a
suspended gold nanoparticle, also via thiol group. By using
different protect group at the two ends, the direction of the PMPH
diblock molecule on the surface can be precisely controlled. The
current-voltage characteristics were measured via scanning
tunneling spectroscopy (STS).  Without protonation, an average
rectification ratio of 7.4 was observed, with the preferential
current direction from the diphenyl block to the dipyrimidinyl
block.  After using perchloric acid in a methanol/tetrahydrofuran
mixture and sodium ethoxide in methanol to protnate and
deprotonate the nitrogen atom in dipyrimidinyl group, a reversible
change in the rectifying direction was observed. The protonated
diblock molecular wire gave an average inverse of the
rectification ratio as 9.2, with current preferring to flow from
dipyrimidinyl side to diphenyl side.

Unfortunately, the microscopic mechanism of this interesting
experimental observation of rectification inversion remains unclear.
Moreles \emph{et al.} \cite{Morales0556} tried to suggest a simple
model to rationalize the rectification inversion. In their model,
before protonation, the intrinsic dipole moment of the diblock
molecule will induce local vacuum level shift and make the highest
occupied molecular orbital (HOMO) closer to electrode Fermi energy.
After the molecule is protonated, the positive charge centered on
the nitrogen reverses the direction of the dipole moment, which
makes the lowest unoccupied molecular orbital (LUMO) becomes closer
to the electrode Fermi energy. But we notice that the relative
position of HOMO/LUMO comparing to the electrode Fermi level is not
directly related to rectifying effect. An orbital approaching Fermi
energy will typically give conductance enhancement, as we can see
later, but rectification should be determined by the bias voltage
response of the relevant orbitals.\cite{taylor0201, li0693, li0616}

In this paper, we report a comprehensive first principles
transport study on PMPH diblock molecular wire. Non-equilibrium
Green's function (NEGF) technique combined with density functional
theory (DFT) is used, which has been widely used in molecular
electronics and successfully applied to molecular rectification
analysis. \cite{taylor0201, li0693, li0616} In the rest part of
this paper, computational details are given in section II. In
section III, we discuss the protonation effect on transport
properties of PMPH molecule wire, with standard hollow site S-Au
anchoring. Different anchoring models are discussed in section IV.
At last, we conclude in section IV.

\section{Computational Methods}

The electronic structures are described with the implementation of
DFT in SIESTA program, \cite{Soler0245} which solves the Kohn-Sham
equation with numerical atomic basis sets. Double-$\zeta$ with
polarization (DZP) basis set is chosen for all atoms except Au,
for which single-$\zeta$ with polarization (SZP) is used. Our test
calculations indicate that using SZP basis set for Au does not
effect the accuracy of our calculations. Core electrons are
modeled with Troullier-Martins pseudopotentials.\cite{Troullier91}
Perdew-Zunger local density approximation \cite{Perdew8148} is
used to describe the exchange-correlation potential.

The electronic transport properties are calculated with NEGF
technique using ATK package, \cite{Brandbyge02, atk} in which the
molecular wire junction is divided into three regions, left
electrode, contact region, and right electrode. The contact region
typically includes parts of  the physical electrodes where the
screening effects take place, to ensure that the charge
distributions in the left and right electrode region correspond to
the bulk phases of the same material. The semi-infinite electrodes
are calculated separately to obtain the bulk self-energy. The
stacking order of the electrode atomic layer is chosen with
inversion symmetry for left and right electrode, which makes the
supercell  not periodic along the current direction. Therefore,
the electrostatic potential is calculated with multigrid method
instead of fast Fourier transformation. A (2 $\times$ 2) and (4
$\times$ 4) $k$-point grid on the $x$-$y$ plane is used for
self-consistent calculation and transmission coeffecients
evaluation respectively.

\section{Protonation Effect on Transport Properties}

\subsection{geometrical model of molecular junctions}

The molecular junction is modeled by sandwiching the diblock
molecule between two Au(111)-(3$\times$3) surfaces with thiol
anchoring group. Before exploring anchoring geometry effects in
the next section, we limit ourselves to the most popular hollow
site adsorption first. Because the PMPH molecule is embeded in
dodecanethiol self-assembly monolayer (SAM) in experiment,
\cite{Morales0556} the lower pyrimidinyl ring should highly
unlikely be protonated. So, we only consider protonaton for one or
both of the two nitragon atoms within the top pyrimidinyl ring.
The resulted molecular junctions are called monoprotonated PMPH
(MP-PMPH) and diprotonated PMPH (DP-PMPH).

\begin{figure}
\includegraphics[keepaspectratio,totalheight=12cm]{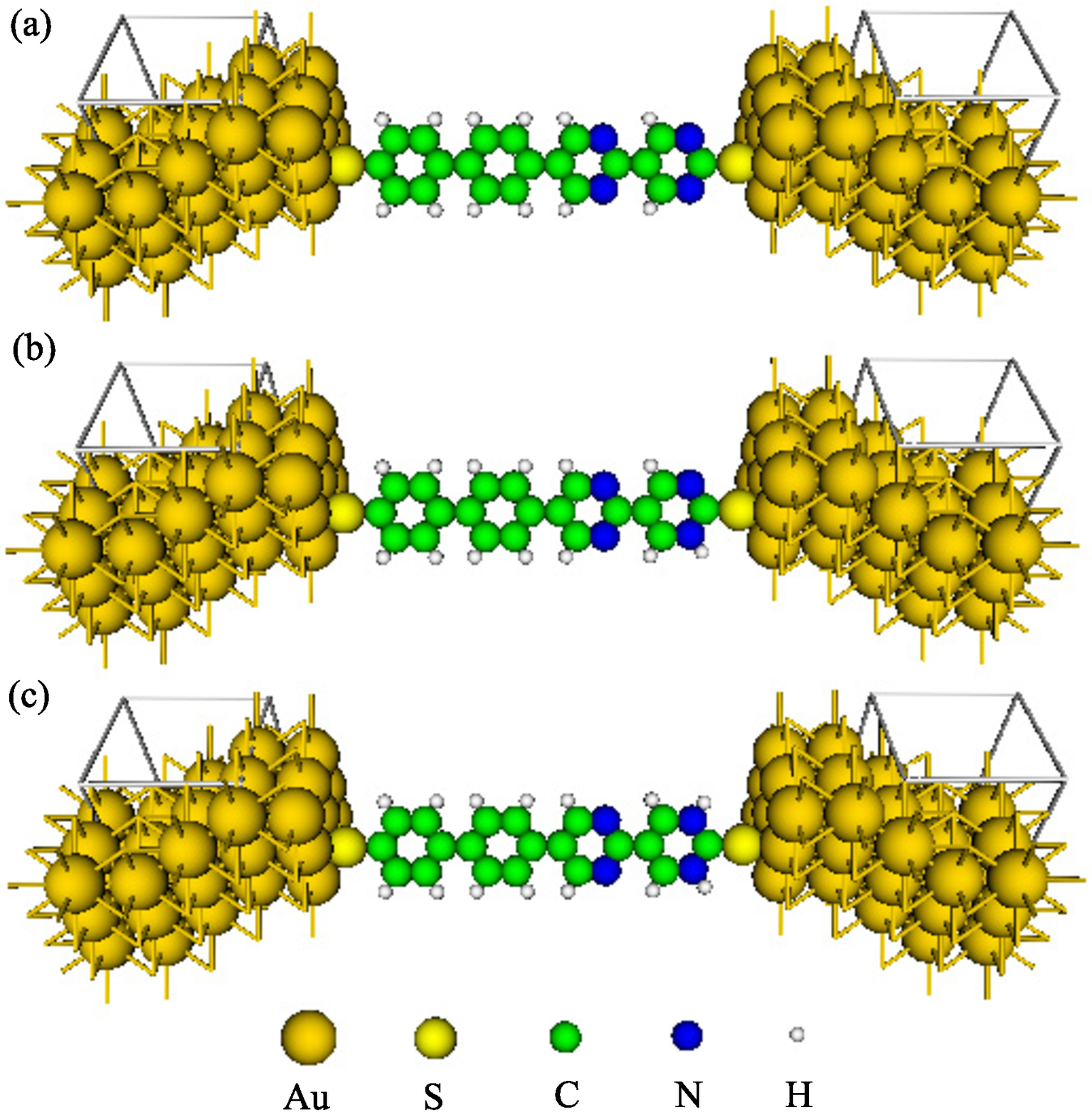}
\caption{The relaxed geometries of Au-molecule-Au junctions. (a)
PMPH, (b) MP-PMPH, and (c) DP-PMPH. } \label{fig:geo}
\end{figure}

Fig. \ref{fig:geo} shows the optimized geometries for the three
molecular junctions with and without protonation. Only one unit
cell of the semi-infinite left/right electrode, which contains
three Au layers, is plotted. The dangling bonds shown in the
figure indicate the periodic boundary condition (PBC) in $x$-$y$
plane and semi-PBC for electrodes in $z$ direction. In the contact
region, two Au layers at both left and right sides are included,
of which the most left/right layer is constrained to its
theoretical bulk geometry to match the structure of the Au(111)
surface. The rest of the contact region is fully optimized. The
distance between the left and right electrodes is determined by a
serial of optimization calculations with different fixed
electrode-electrode distances.

\subsection{transmission and current-voltage characteristics}

\begin{figure}
\includegraphics[keepaspectratio,totalheight=10cm]{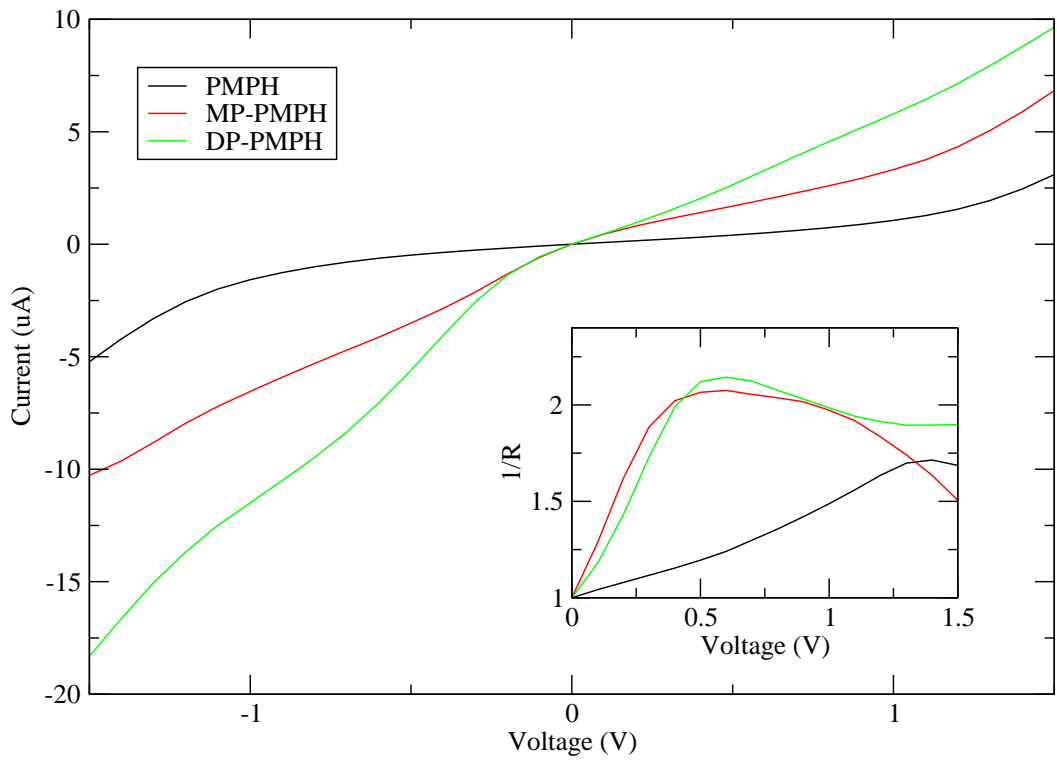}
\caption{Current-voltage curves for PMPH, MP-PMPH, and DP-PMPH
molecular junctions. Inset: corresponding inverse of rectification
ratios.} \label{fig:iv}
\end{figure}

Based on the optimized geometries, the current-voltage ($I$-$V$)
curves for PMPH, MP-PMPH, and DP-PMPH junctions have been
calculated self-consistently, as shown in Fig. \ref{fig:iv}.
Almost symmetric currents through PMPH molecular wire are obtained
for positive and negative bias voltages, especially at low
magnitudes of bias voltage. However, obvious asymmetric currents
are observed for MP-PMPH and DP-PMPH, for which negative bias
leads to larger current. The asymmetry of the $I$-$V$ curves can
be measured by rectification ratio ($R$), which is define by
$R(V)=|I(V)/I(-V)|$.  In the inset of Fig. \ref{fig:iv}, we plot
inverse of $R$ as a function of bias voltage magnitude for all
these three junctions. $R$ approaches to 1 at low bias voltage for
all these three junctions, which means that the slope of I-V curve
does not change abruptly at zero bias. Although the rectification
ratios for MP-PMPH and DP-PMPH are distinguished with that of PMPH
by their much larger values, the rectifying directions are same
for all these three junctions, with current preferring to flow
from dipyrimidinyl side to diphenyl side. Besides rectification
effect, protonation also leads to conductance enhancement.  The
conductance of MP-PMPH and DP-PMPH is about a factor of 2 and 4
respectively larger than that of PMPH. We notice that the
calculated currents are an order of 5 larger than the experimental
values. It may be partly caused by the high vacuum barrier between
gold nanoparticle and STM tip electron should overcome in
experiment. \cite{Morales0556}

To understand the different electron transport properties for PMPH
molecular wire before and after protonation, we plot the
transmission curves for PMPH, MP-PMPH, and DP-PMPH at zero bias
voltage in Fig. \ref{fig:trans}. The main common feature of the
three transmission curves is that there are four broad peaks above
Fermi level up to 4 eV. Below the fermi energy, above -2.5 eV,
there are two transmission peaks. The one higher in energy is
however very low in transmission. For PMPH, the four
above-Fermi-level peaks are almost perfect resonances with
transmission possibility close to one. However, for MP-PMPH and
DP-PMPH, the first of these four peaks is much low than one. As we
can see, the conductance at zero bias voltage is mainly determined
by the tail of this peak. We name it peak $A$ thereafter. Although
the hight of peak $A$ is much smaller for MP-PMPH and DP-PMPH than
that for PMPH, the Fermi energy is farer away from peak $A$ for
PMPH. As a result, the conductance of PMPH is smaller than those
of MP-PMPH and DP-PMPH. From the transmission curves, we can
expect that MP-PMPH and DP-PMPH have similar conductance at zero
bias voltage, which is exactly what we observed in Fig.
\ref{fig:iv}.

To consider the finite bias voltage transport and rectification
behavior,  the voltage response of transmission spectrum should be
studied. In Fig. \ref{fig:trans}, we also plot the transmission
curves at a bias voltage of $\pm$1.0 V. Because the heights and
positions of peak $A$ is more strongly affected by the bias
voltage in MP-PMPH and DP-PMPH junctions than that in PMPH
junction, the latter gives much smaller rectification effect. We
notice that for MP-PMPH, there is a sharp transmission peak
overlapping with peak $A$ at zero bias voltage, and closing to
peak $A$ at -1.0 bias voltage. At negative bias, this sharp peak
even give near perfect transmission, but since the peak is narrow
and relatively far away from the Fermi level, it does not really
contribute to the low bias conductance of MP-PMPH.

\subsection{MPSH orbital analysis}

\begin{figure}
\includegraphics[keepaspectratio,totalheight=11cm]{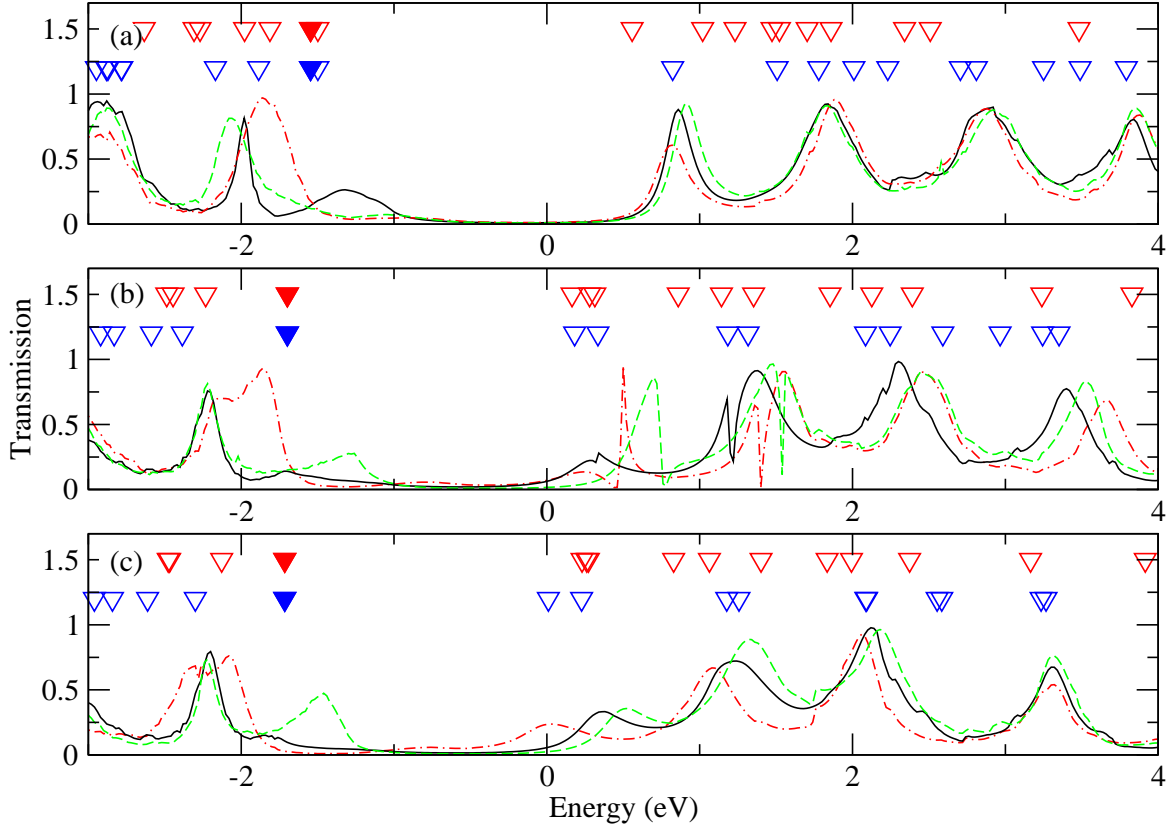}
\caption{Transmission curves at zero (black solid), -1.0 V (red
dash-dot), and 1.0 V (green dash) bias voltage for (a) PMPH, (b)
MP-PMPH, and (c) DP-PMPH molecular junctions. Energies of
molecular orbitals and MPSH orbitals at zero bias voltage are
indicated by upside down triangles. The averaged electrode Fermi
energy is set to zero, and there is a rigid energy shift for
molecular orbitals.} \label{fig:trans}
\end{figure}

The transmission peaks can be considered as resonant transmission
of the electron through renormalized molecule orbitals, which can
be obtained by diagonalizing molecular projected self-consistent
Hamiltonian (MPSH). Both the energies of orbitals of free molecule
(with two thiol groups at two ends) and MPSH orbitals at zero bias
voltage are marked by upside-down triangles in Fig.
\ref{fig:trans}. The first row corresponds to molecular orbitals,
and the second row is MPSH eigenvalues. By comparing the energies,
we notice that almost all transmission peaks are mainly
contributed by a MPSH orbital. At the same time, the MPSH orbitals
are closely related to molecular orbitals. Typically, we can find
a one-to-one map for molecular orbitals and MPSH orbitals by
comparing their real space distribution. \cite{li0693, li0616,
wu0512} As shown in Fig. \ref{fig:trans}, their energies are also
comparable after a rigid shift. The amount of the rigid energy
shift for molecular orbitals is chosen to align an arbitrarily
chosen molecular orbital with its corresponding MPSH orbital, as
indicated by solid triangles in Fig. \ref{fig:trans}. There are
several reasons why an energy shift is necessary. The real part of
the electrode self energy may shift the MPSH energy, and, as
suggested by Morales et al., \cite{Morales0556} the dipole moment
of the molecule may also shift local vacuum level.

\begin{figure}
\includegraphics[keepaspectratio,totalheight=10cm]{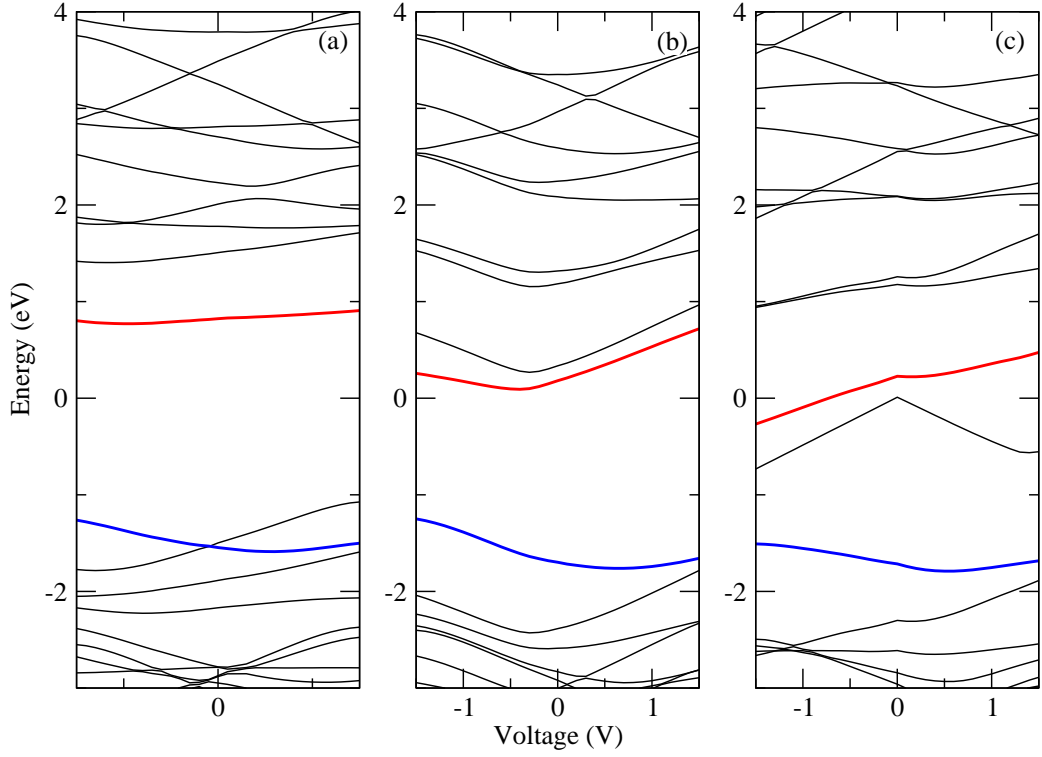}
\caption{Voltage response of MPSH eigenvalues for (a) PMPH, (b)
MP-PMPH, and (c) DP-PMPH molecular junctions. There are two energy
band plotted bold for each junction. The blue one comes from HOMO
of PMPH molecule, and is chosen to align molecular orbital and
MPSH orbital in Fig. \ref{fig:trans}. The similar energy and bias
voltage response of this MPSH orbital for all these three
junctions indicate it is a good choice for energy alignment
purpose. The red one comes from LUMO of PMPH molecule, and it
corresponds to the transmission peak $A$, which determines the low
bias transport properties.} \label{fig:eig}
\end{figure}

\begin{figure}
\includegraphics[keepaspectratio,totalheight=6cm]{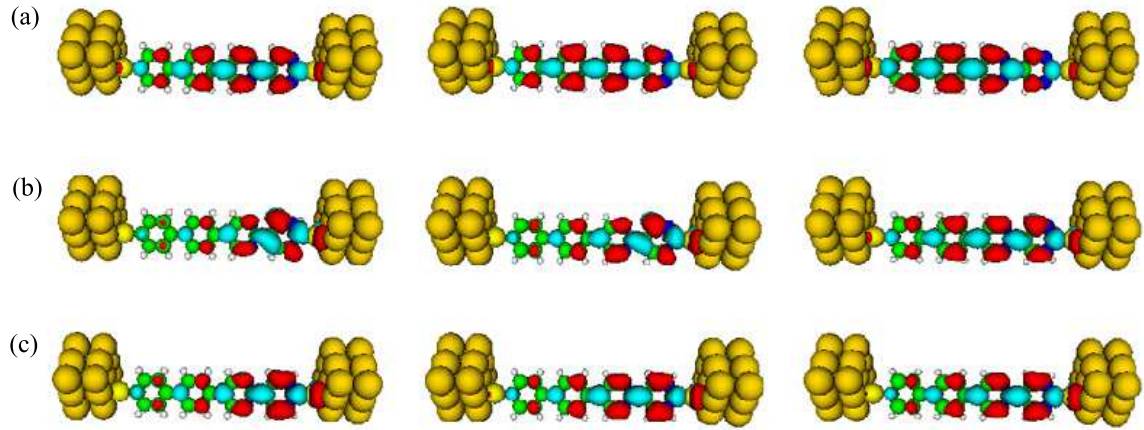}
\caption{MPSH orbitals corresponding to the transmission peak $A$,
which determines the low bias transport properties, for (a) PMPH,
(b) MP-PMPH, and (c) DP-PMPH molecule. From left to right, MPSH
orbitals at -1.0, 0, and 1.0 bias voltage respectively.  }
\label{fig:mpsh}
\end{figure}

Voltage response of transmission can also be analyzed by MPSH
orbitals. Bias voltage may change both the position and the shape of
a transmission peak. The peak position generally will follow the
energy of corresponding MPSH orbital. In Fig. \ref{fig:eig}, we plot
MPSH energies versus bias voltage, which can also be roughly
considered as transmission peak positions versus bias voltage. The
line shape and height of transmission peak will be determined by the
MPSH orbital real space distribution (transmission channel). Well
delocalized transmission channel will give almost perfect
transmission (close to one), and strong electrode-molecule coupling
results broad transmission peak. For all three junctions studied
here, according to the position of the Fermi energy, the low bias
transport is mainly determined by peak $A$, the first
above-Fermi-level peak. The real space distribution of the
corresponding MPSH orbital of peak $A$ is plotted in Fig.
\ref{fig:mpsh}.

For PMPH, peak $A$ mainly comes from MPSH LUMO. As shown in Fig.
\ref{fig:eig} and \ref{fig:mpsh}, the energy of MPSH LUMO is only
slightly affected by the bias voltage, and the real space
distribution is also not very sensitive to bias voltage. As a
result, the position of transmission peak $A$ is almost
independent with the bias voltage, and peak width and height are
only slightly changed. For MP-PMPH, the corresponding MPSH orbital
is relatively insensitive to negative bias voltage, but become
much more symmetric and delocalized at positive bias voltage.
Therefore, its peak $A$ is much higher at positive bias voltage.
However, the MPSH orbital also upshift much at positive bias
voltage, which makes peak $A$ far from Fermi energy. Finally, we
get smaller conductance at positive bias voltage. The sharp peak
overlapped with peak $A$ at zero bias comes from the next MPSH
orbital, which is mainly distributed within bipyrimidinyl group,
and thus gives only a low transmission possibility.

It is interesting to note that the transmission response to bias
voltage for peak $A$ behaviors in the same way for all these three
junctions. At negative bias, the peak shift to lower energy, and
its magnitude decrease. At positive bias, we get both higher peak
position and height. At a first look, it is strange,  since the
molecules are oppositely polarized. However, if we look at the
corresponding MPSH orbitals for peak $A$ in Fig. \ref{fig:mpsh},
we find that they are all polarized at the same direction. They
are all distributed in the dipyrimidinyl group more than in
diphenyl group.  Of course, the magnitude of polarization is very
different. The MPSH orbital for PMPH junction is almost
unpolarized, so it is well delocalized within the whole molecule,
and gives very high transmission. For MP-PMPH and DP-PMPH, the
MPSH orbital is highly polarized even at zero bias, therefore peak
$A$ is much lower than one, and the peak position is more strongly
dependent on the bias voltages. As a result, MP-PMPH and DP-PMPH
gives much larger rectification coefficient at small bias voltage.
Therefore, to understand the transport properties, we should look
into the electronic structure in more details, not only the
polarization of the whole molecule, but also the polarization and
its voltage response of the relevant orbitals which are close to
Fermi level.

\section{Effects of molecular anchoring model}

\begin{figure}
\includegraphics[keepaspectratio,totalheight=16cm]{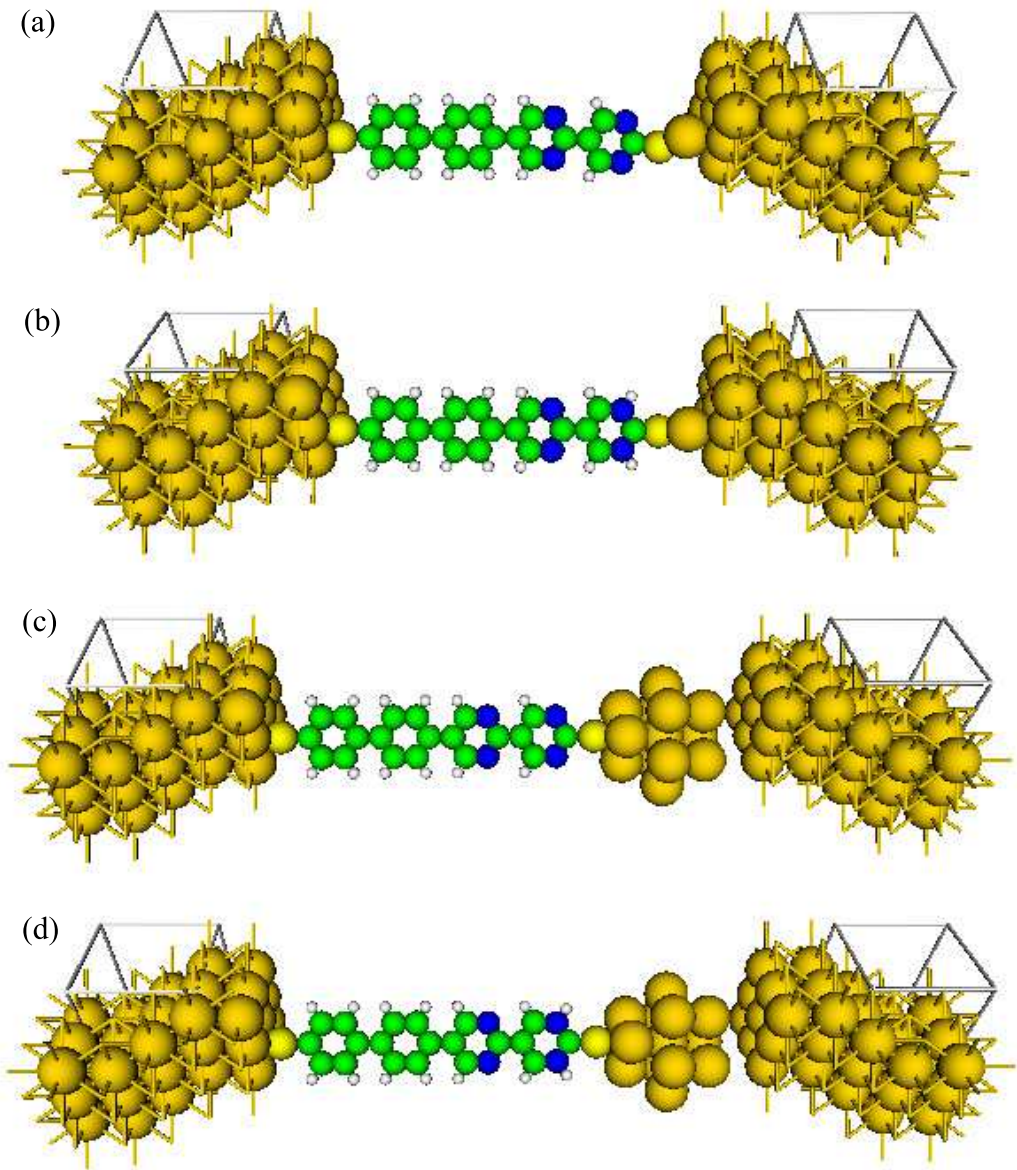}
\caption{Geometry of Au-molecule-Au junctions. (a)  PMPH-A, (b)
DP-PMPH-A junction, (c) PMPH-C, and (d) DP-PMPH-C. }
\label{fig:geo2}
\end{figure}

Until now, the thiol anchoring group is connected to gold surface
through the fcc hollow site. This is a standard model, but it may
not be adequate to describe the experimental setup by Morales et
al., \cite{Morales0556} where a gold nanoparticle is suspended.
The surface of a nanoparticle may be far away from a clean (111)
surface. On the other hand, the nanoparticle may address some
stress to the diblock molecule, and the electrode-electrode
distance may be much different from its equilibrium value. Another
big issue is that the STM tip typically is not contact with the Au
nanopariticle in STS measurements. Based on these concerns, we
construct two more geometrical models for PMPH and DP-PMPH, namely
apex model and cluster model. In apex model, we add an apex Au
atom on Au (111) surface at the bipyrimidinyl side, and obtain
junctions PMPH-A and DP-PMPH-A. In these two junctions, the
distance of the two electrode is determined by an optimization
with electrode represented by a Au$_4$ cluster. The cluster model
is constructed with the diblock molecule connecting to a sphere
shaped Au$_{13}$ cluster, which is separated with the Au(111)
surface by 4 \AA. The resulted two junctions are called PMPH-C and
DP-PMPH-C. The geometrical structures of these four junctions are
plotted in Fig. \ref{fig:geo2}.

\begin{figure}
\includegraphics[keepaspectratio,totalheight=10cm]{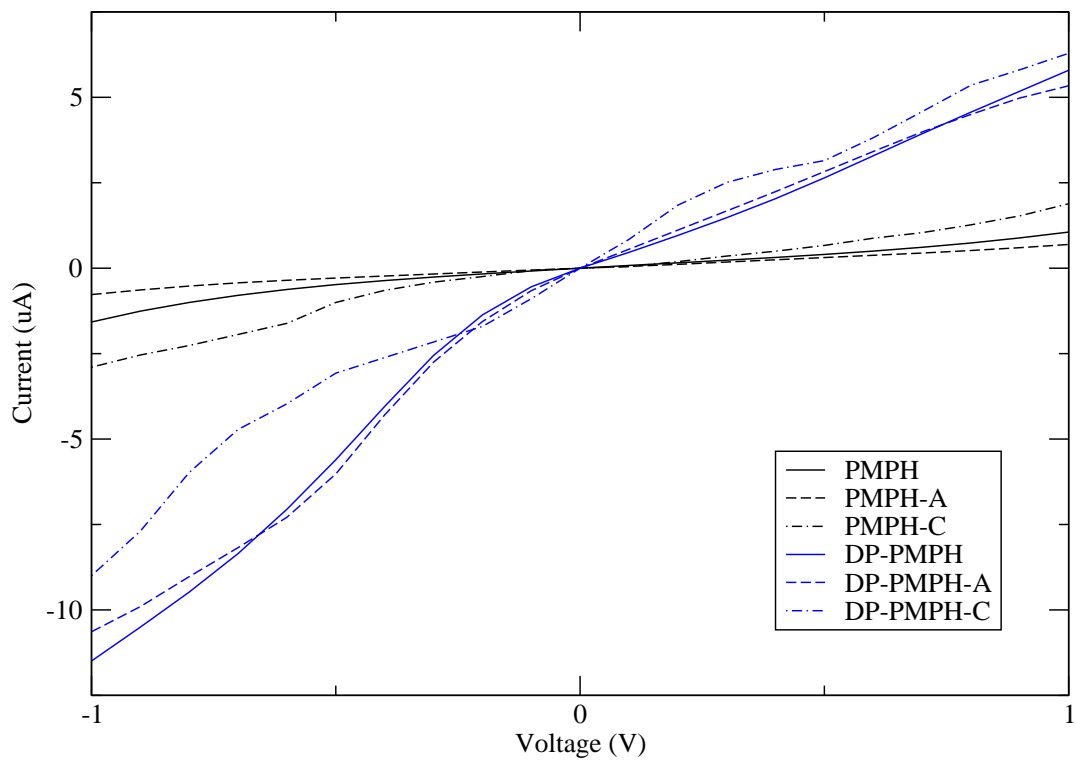}
\caption{Current-voltage curves for PMPH, PMPH-A, PMPH-C, DP-PMPH,
DP-PMPH-A, and DP-PMPH-C molecular junctions.} \label{fig:iv2}
\end{figure}

Despite the difference of anchoring geometries, the calculated
current-voltage curves, as shown in Fig. \ref{fig:iv2}, however,
are qualitatively same for standard, apex, and cluster model.
Especially for apex and standard model, the difference of I-V
curves is very small. The cluster model gives relatively larger
difference, but still no rectification inversion observed. This
result indicates that the failure to reproduce of the experimental
rectification inversion is not a result of unrealistic anchoring
model.  It is interesting to notice that, with a 4 \AA\, gap, the
current through PMPH-C junction is larger than that through PMPH
junction. For DP-PMPH-C, current increases comparing to that of
DP-PMPH at positive bias voltages, and decreases at negative bias
voltages.

\begin{figure}
\includegraphics[keepaspectratio,totalheight=10cm]{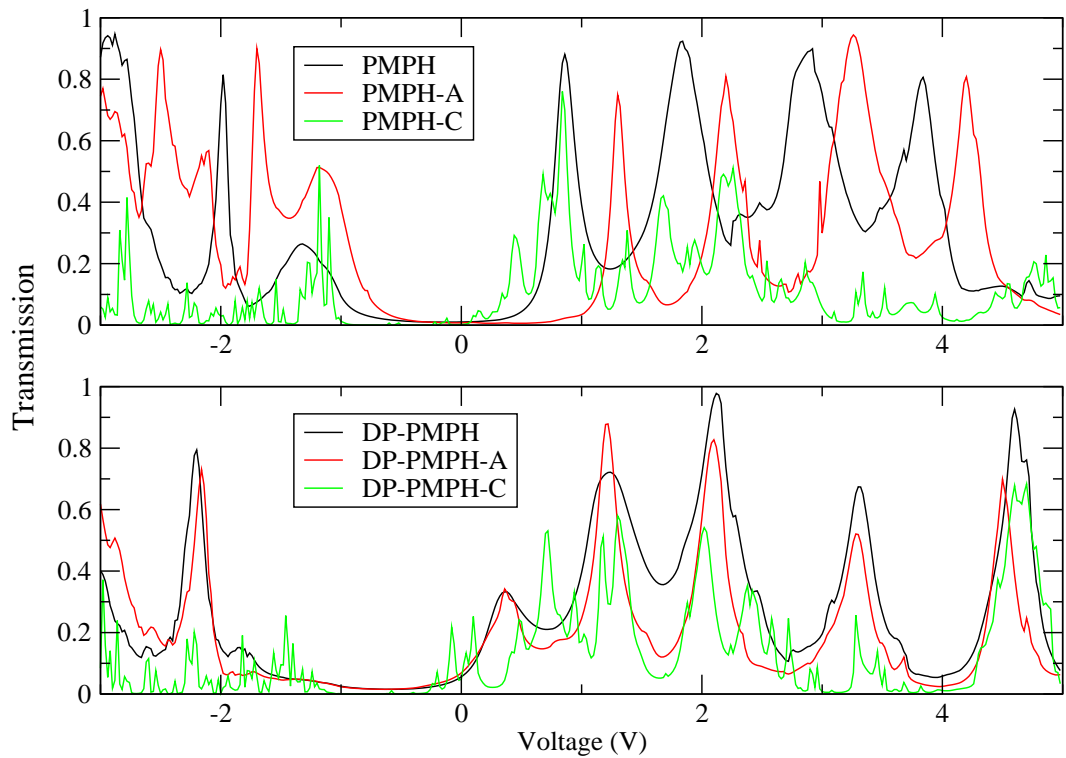}
\caption{Transmission curves at zero bias voltage for PMPH,
PMPH-A, and PMPH-C, DP-PMPH, DP-PMPH-A, and DP-PMPH-C molecular
junctions.  } \label{fig:trans2}
\end{figure}

The conductance behavior can be understood from transmission
spectra. In Fig. \ref{fig:trans2}, we plot the transmission spectra
for these junctions. For apex model, as we expected, the
transmission spectra are more or less the same as those for hollow
model, except that there is a shift for PMPH-A comparing to PMPH.
For cluster model, transmission characters are much more complicated
than standard model, because the electronic structure of the gold
cluster manifest itself in, as it does in dithiocarboxylate
anchoring group.\cite{li0616} The interesting conductance
enhancement of cluster model comes from some new small peaks very
close to Fermi energy. Therefore, for off-resonance transport, it
not obvious to get small conductance with weaker electrode-molecule
coupling. \cite{comment} At the same time, the richer transmission
feature brought by Au cluster makes the I-V curve less smooth for
the cluster model junctions.

\section{Conclusions}

Transport properties of PMPH diblock molecular wire with and without
protonation are studied theoretically by combining NEGF and DFT.
Protonation is found to be able to enhance conductance and
rectification, but no rectifying direction inversion is found in
this study. Anchoring geometry effect is carefully checked and it
turns out that it is not relevant in this issue. Our calculations
indicate that the rectification inversion observed in experiment may
not be an intrinsic molecular property related to protonation, and
more sophisticated theory should be developed to explain this
experiment. There may be two important things missed in this study.
One is the interaction between molecular wire and environmental
dodecane and solvent molecules, and the other is the electron
correlation beyond the NEGF+DFT level of theory.

\section*{acknowledgments}
The author is grateful to D. S. Kosov for helpful discussion.

\clearpage

\clearpage

\clearpage

\clearpage

\clearpage

\clearpage

\clearpage

\clearpage


\begin{thebibliography}{99}
\bibitem{comment}We also want to emphasize that this conductance
enhancement of cluster model comparing to standard model may be an
artificial result at the NEGF+DFT level of theory, which does not
consider electron correlation very well.
\end{thebibliography}
\end{document}